\documentclass[10pt,letterpaper]{article}

\usepackage{ccn}
\usepackage{pslatex}
\usepackage{apacite}
\usepackage{graphicx}
\usepackage{caption}
\usepackage{blindtext}
\usepackage{amsmath}
\usepackage{tabularx}
\usepackage{fancyhdr}
\fancypagestyle{firstpage}{%

  \lfoot{\vspace{5mm}Accepted at the 2022 Conference on Cognitive Computational Neuroscience (CCN 2022)}
}

\title{Bayesian Modeling of Language-Evoked Event-Related Potentials}
 
\author{{\large \bf Davide Turco (davide.turco@bristol.ac.uk)}\\\large \bf Conor Houghton (conor.houghton@bristol.ac.uk)\\
Department of Computer Science, University of Bristol, Bristol, BS8 1UB, UK}

\begin{document}

\maketitle
\thispagestyle{firstpage}

\section{Abstract}
{
\bf
Bayesian hierarchical models are well-suited to analyzing the often noisy data from electroencephalography experiments in cognitive neuroscience: these models provide an intuitive framework to account for structures and correlations in the data, and they allow a straightforward handling of uncertainty. In a typical neurolinguistic experiment, event-related potentials show only very small effect sizes and frequentist approaches to data analysis fail to establish the significance of some of these effects. Here, we present a Bayesian approach to analyzing event-related potentials using as an example data from an experiment which relates word surprisal and neural response. Our model is able to estimate the effect of word surprisal on most components of the event-related potential and provides a richer description of the data. The Bayesian framework also allows easier comparison between estimates based on surprisal values calculated using different language models.
}
\begin{quote}
\small
\textbf{Keywords:} 
Natural language processing; Neurolinguistics; Bayesian modeling; ERP; EEG; Hierarchical models
\end{quote}

\section{Introduction}
Electroencephalography (EEG) is a non-invasive technique used to investigate the cognitive processes underlying language comprehension. In neurolinguistics, it is common to work with event-related potentials (ERP), which are time-locked brain responses measured for a specific stimulus, such as a word. As an example, in a study of the response to the information conveyed by words \cite{DAMBACHER200689}, it was observed that particularly surprising words affect the N400 component, the negative potential deflection peaking around 400 ms post-stimulus onset. 

ERP data are usually analyzed using frequentist methods, such as traditional \cite{Broderick2018} or mixed-effect \cite{Frank2015} linear models, or using neural networks \cite{Schwartz2019}. Although these approaches are well-established and efficient, it is difficult to include prior beliefs for effect magnitudes or correlation structures, uncertainty is not explicitly quantified and they tie statistical modeling to an often intricate or misleading hypothesis-testing framework.

Here, we propose a Bayesian analysis of ERP data, using the data described in \citeA{Frank2015} as an example.  We show that a full Bayesian treatment not only replicates results from traditional statistics in a more elegant and data-efficient way, but it also supports the inclusion of insights from neurolinguistics and the structure of the experiment itself. Previous work employing Bayesian approaches focused on the relationship between cloze probability and just one component, the N400 \cite{Nicenboim2022}.  

\section{Methodology}
In this example, we used openly available data previously described in \citeA{Frank2015}. ERPs were estimated from EEG recordings from 24 participants reading 205 sentences, corresponding to 1931 words; these stimulus sentences are orthographically and grammatically correct. 
Words were tagged for their part-of-speech (POS) using the Universal Tagset \cite{petrov-etal-2012-universal} and divided in content (nouns, verbs, adjectives, adverbs) and function (any other POS tag) words. The six ERP components investigated in this study are ELA\underline{N}, LA\underline{N}, \underline{P}600, generally considered markers of syntactic processing, and EPN\underline{P}, PN\underline{P}, \underline{N}400, well-known markers of semantic processing. \underline{N} or \underline{P} indicate negativity or positivity with respect to the pre-stimulus onset baseline.

Surprisal to the information conveyed by a word $w_t$ given the preceding words $w_{1,\dots,t-1}$ is
\begin{equation}
\text{surprisal}(w_t)=-\log P(w_t|w_{1,\dots,t-1}).
\end{equation}
The surprisal values can be estimated using any probabilistic language model (LM) for $P(w_t|w_{1,\dots,t-1})$. We used surprisal values from $n$-grams, as computed by \citeA{Frank2015} with $n=3$; we also calculated surprisal estimates from an LSTM and a transformer. The LSTM \cite{Hochreiter1997} had two hidden layers and was trained on the WikiText-2 Corpus \cite{Merity2016} with an LM objective. The transformer was the causal LM GPT-2 \cite{Radford2018}, pre-trained by processing input left-to-right.

\begin{figure*}[ht]
\begin{tabularx}{.99\linewidth}{ 
  X X X  }
\textbf{A}&\textbf{B}&\textbf{C}\\[-3pt]
\includegraphics[width=0.32\textwidth]{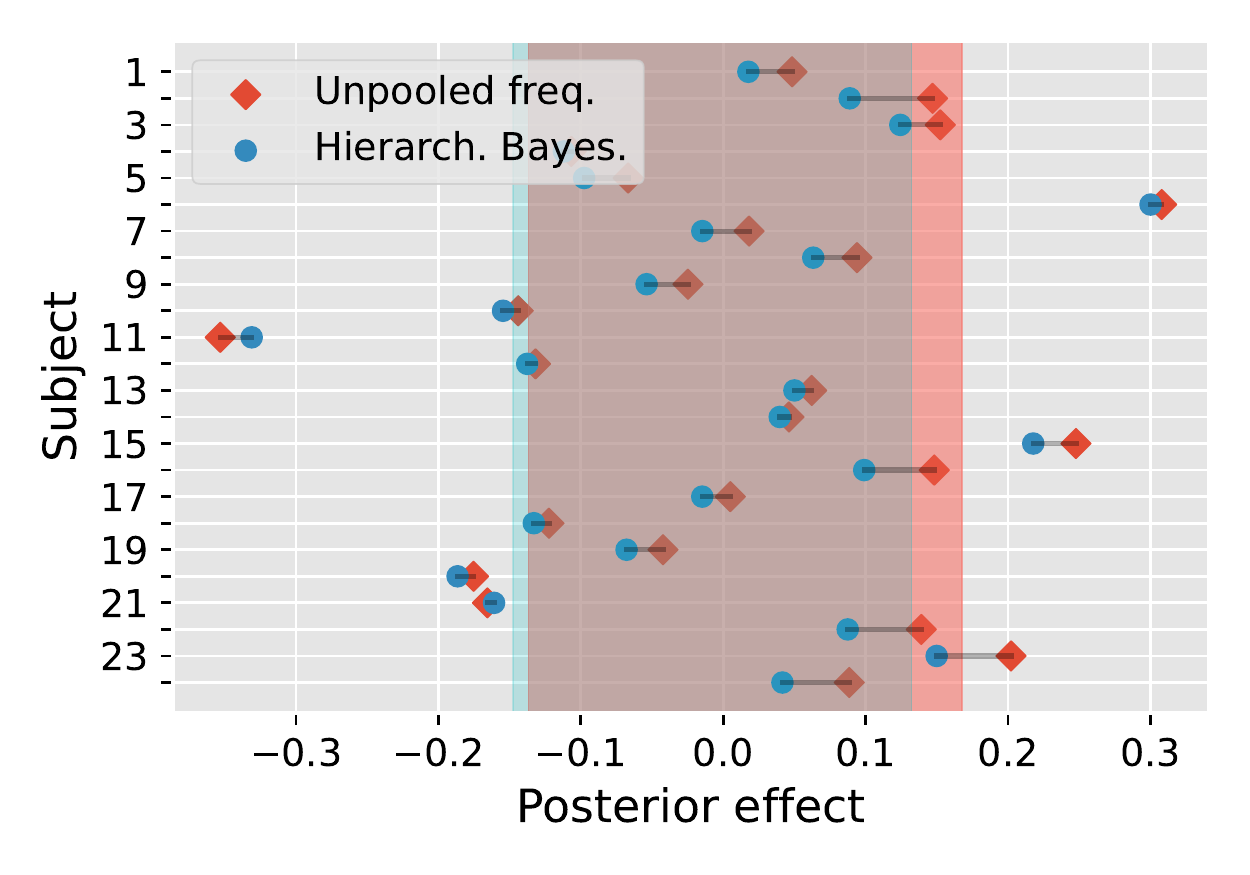}&
\includegraphics[width=0.32\textwidth]{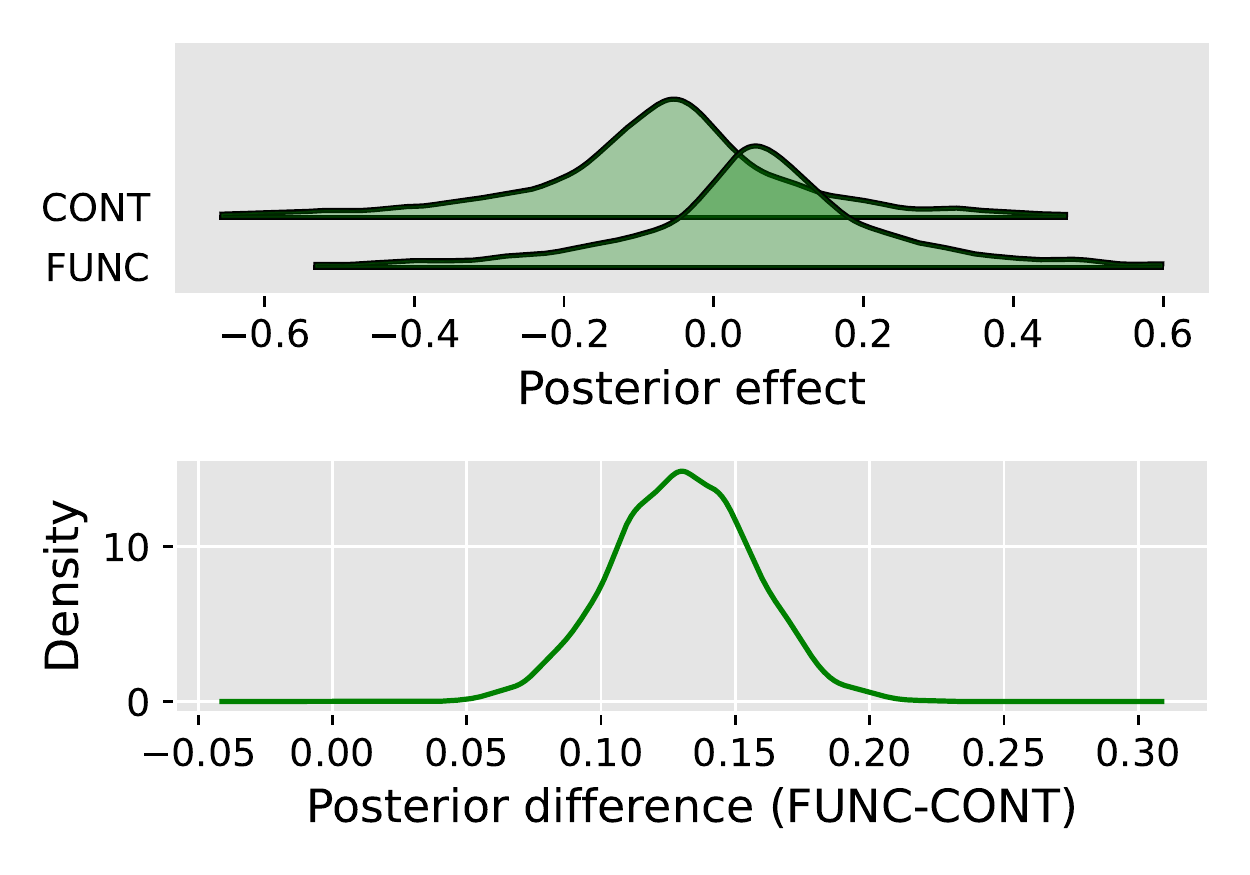}&
\includegraphics[width=0.32\textwidth]{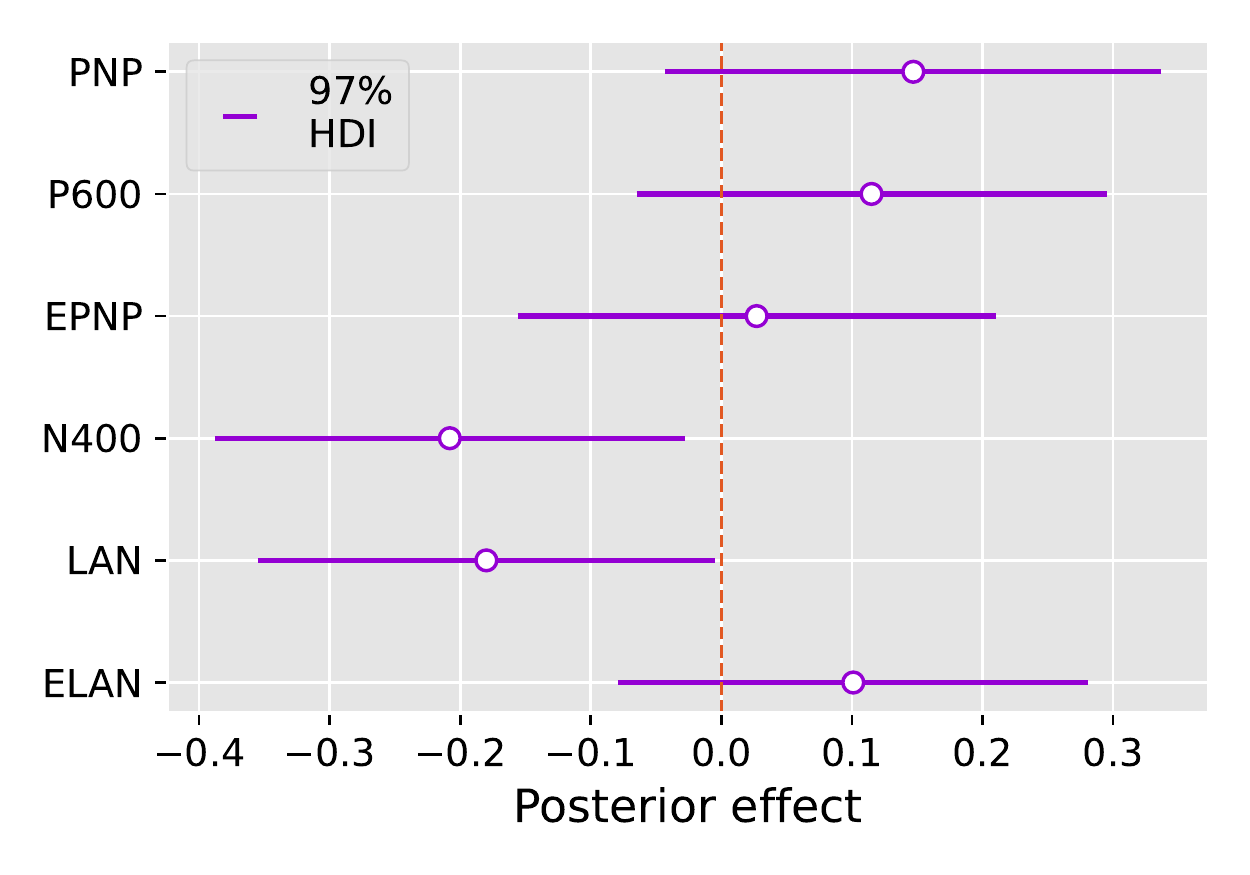}
\end{tabularx}
\caption{Main insights from the posterior distributions for $n$-gram surprisal values. Similar results are obtained with the LSTM or GPT-2. \textbf{A} Our hierarchical Bayesian model shrinks by-subject effects closer to the mean, leading to a smaller standard deviation (shaded area) than a traditional unpooled linear model. \textbf{B} Content (CONT) and function (FUNC) words contribute differently to the prediction; the posterior probability of their difference is non-zero. \textbf{C} All the ERP components are modeled with similar uncertainties, but EPNP clearly favors no effect.}
\label{fig:results}
\end{figure*}
We modeled ERP data with a hierarchical Bayesian model. Unlike \citeA{Frank2015}, data from all the ERP components were analyzed together in a single model. The analysis was treated as a regression problem, with standardized surprisal $s$ as the predictor and ERP amplitude $r$ as the response variable. The likelihood was given a normal distribution, centered at a mean estimated using a varying-coefficient linear model: 
\begin{align}
     r_i&\sim\text{Normal}(\mu_i,\sigma)\\
     \mu_i&=(a_i+a_{p[i]}+a_{w[i]}+a_{e[i]})+(b_i+b_{p[i]}+b_{w[i]}+b_{e[i]})s_i\notag
\end{align}
where $i$ is an item index, $p[i]$, $w[i]$ and $e[i]$ map from item to participant, word type and ERP component and $\sigma$ is the regression error, which has a weakly informative half-Cauchy prior. By-subject, by-word-type and by-ERP-component random effects were given multivariate normal (MVN) priors, with a correlation structure across regression coefficients. An LKJ prior was chosen for their covariance matrices. The same effects could have been included into a frequentist mixed-effect model. However, a Bayesian analysis allows the formulation of priors informed by domain knowledge: for example, the effect means were given tight normal priors centered at zero that reflect the expected small amplitudes of these contributions.

The posterior was sampled with the No-U-Turn Sampler \cite{Hoffman2014}, using the probabilistic library \texttt{PyMC v.4b} \cite{Salvatier2015} in \texttt{Python 3.9}. We sampled four chains for $1000$ samples and used a non-centered parameterization for the MVN priors. 

\section{Results}
The posterior distributions for the model fitted on $n$-gram surprisal values are illustrated in Fig. \ref{fig:results}. One of the immediate effects of using a hierarchical model is that estimates are closer to the population mean than estimates from an unpooled model: this is evident for by-subject effects in Fig. \ref{fig:results}\textbf{A}. Shrinkage is a desirable feature when working with noisy data, as it makes the model more robust to outliers. 

By including by-word-type effects, it is easy to see that there is a difference in the response for content and function words, as shown in Fig. \ref{fig:results}\textbf{B}. This is evident without the need to refit the model on each of the two word types individually, as done by \citeA{Frank2015}. The response to content words is negative, consistent with a N400 effect.

Our model estimates non-zero amplitudes for the N400, LAN and, with less certainty, the PNP, P600 and ELAN components, as illustrated in Fig. \ref{fig:results}\textbf{C}. Our average coefficients for the N400 are comparable to those reported in \citeA{Frank2015}, which are $-0.17$ and $-0.22$ for $n$-grams and a recurrent LM respectively. The polarity of the components is respected by all the models, with the exception of ELAN, which is estimated to be a positive component.
The difficulty in estimating the amplitude of ELAN may be due to the absence of syntactic violations in the stimuli \cite{Friederici2011}. The model fitted on GPT-2 surprisal values gives even more density to ELAN being a positive component, but it favors a near-zero amplitude for P600.

Models fitted on the surprisal values from the three LMs were evaluated using leave-one-out (LOO) cross-validation with Pareto smoothing \cite{Vehtari2017}. The comparison revealed that, taking into account the uncertainty in LOO estimation, no LM's surprisal outperforms the others: this may indicate that even simpler LMs can be used to estimate word information values. Moreover, they all lead to similar results as those illustrated in Fig. \ref{fig:results}. All the models converge satisfactorily, with an upper-bound for R-hat of $1.007$.

\section{Discussion}
We presented a Bayesian approach to modeling ERP data from neurolinguistics experiments, using data reported in \citeA{Frank2015} as motivation. Our model estimates the amplitudes of more ERP components than the original study, that could only predict the N400, and it also describes the data at different levels. Additional experiments, not included here, confirm that our approach can establish the same results with a subset of the subjects or words. 

The main limitation of the outlined approach is the increased computational time: sampling the posterior took around six hours on four CPUs with GPU acceleration. This could be partly improved by better investigating the geometry of the posterior and using alternative parametrizations. Additional prior predictive checks could also be beneficial. In future, we aim to extend the model so that it estimates the event-related EEG directly, rather than individual components which have been calculated by binning and averaging across electrodes: this would allow us to capture information at electrode level. This approach can be extended to other word properties and sources of data, such as magnetoencephalography.

\section{Acknowledgments}
We thank Cian O'Donnell and Sydney Dimmock for useful feedback. DT is funded by an EPSRC Centre of Doctoral Training grant (EP/S022937/1) and CH is a Leverhulme Research Fellow (RF-2021-533). This work was carried out using the HPC facilities of the ACRC, University of Bristol.




\bibliographystyle{apacite}

\setlength{\bibleftmargin}{.125in}
\setlength{\bibindent}{-\bibleftmargin}

\bibliography{ccnbib.bib}

\newpage
\appendix
\onecolumn
\section{Supplementary material\protect\footnote{Not included in the version of the paper accepted at CCN 2022.}}

\subsection{Data and Code Availability}
The EEG and ERP data, as well as the $n$-gram surprisal values, were taken from \citeA{Frank2015} and are openly available\footnote{\url{https://ars.els-cdn.com/content/image/1-s2.0-S0093934X15001182-mmc1.zip}}. The code for performing the Bayesian analysis and reproducing the images on this paper is also available on GitHub\footnote{\url{https://github.com/davideturco/BayesERPs}}.

\subsection{Additional Results}
\begin{figure*}[h]
\begin{tabularx}{.99\linewidth}{ 
  X X X  }
\textbf{A}&\textbf{B}&\textbf{C}\\[-3pt]
\includegraphics[width=0.32\textwidth]{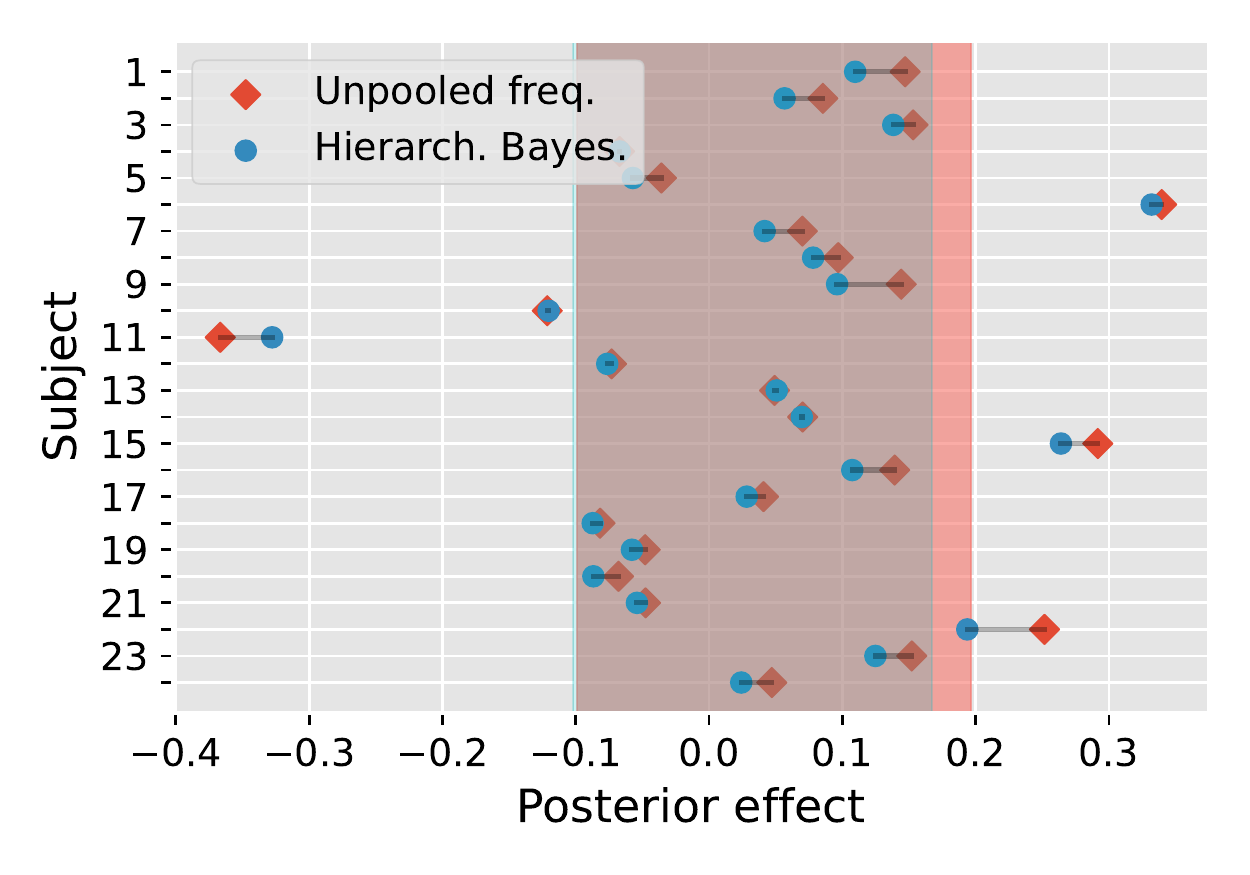}&
\includegraphics[width=0.32\textwidth]{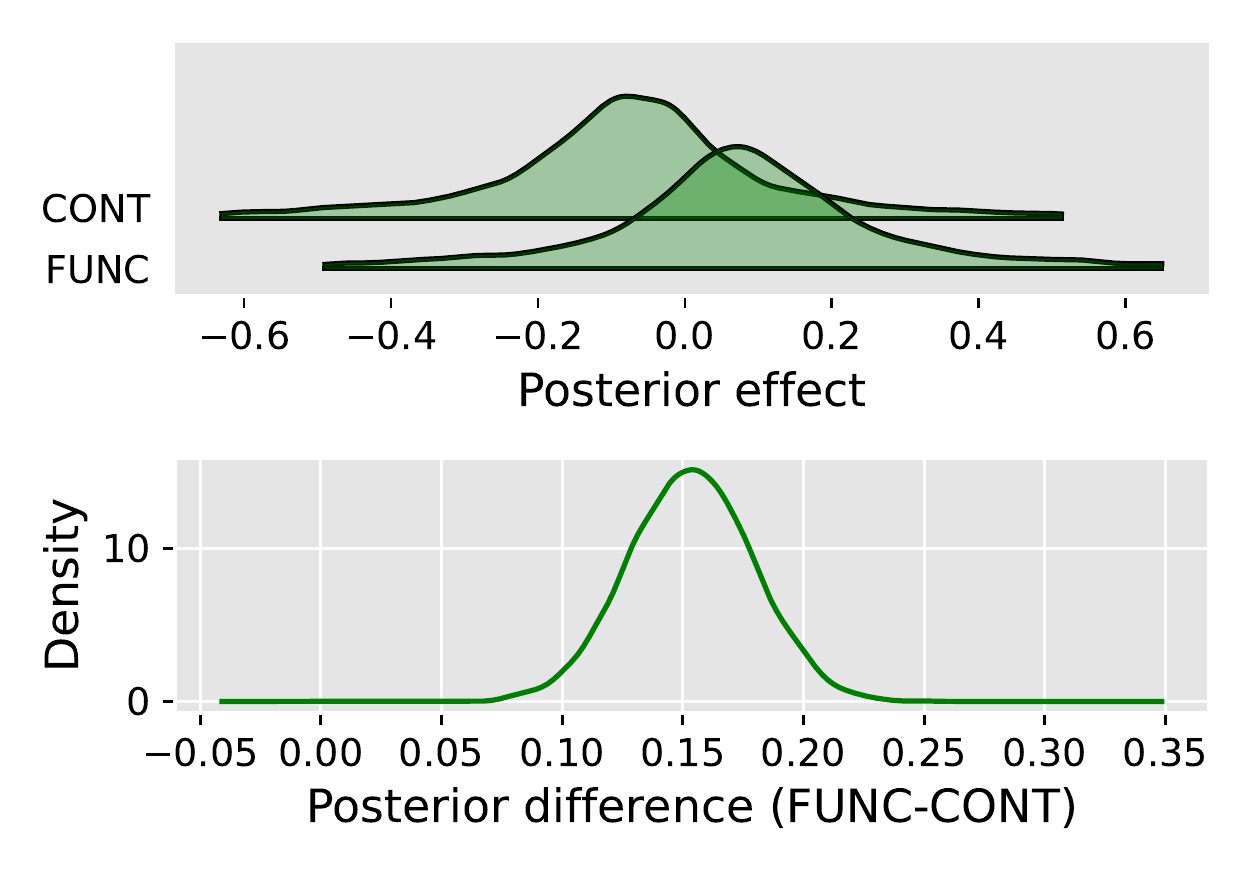}&
\includegraphics[width=0.32\textwidth]{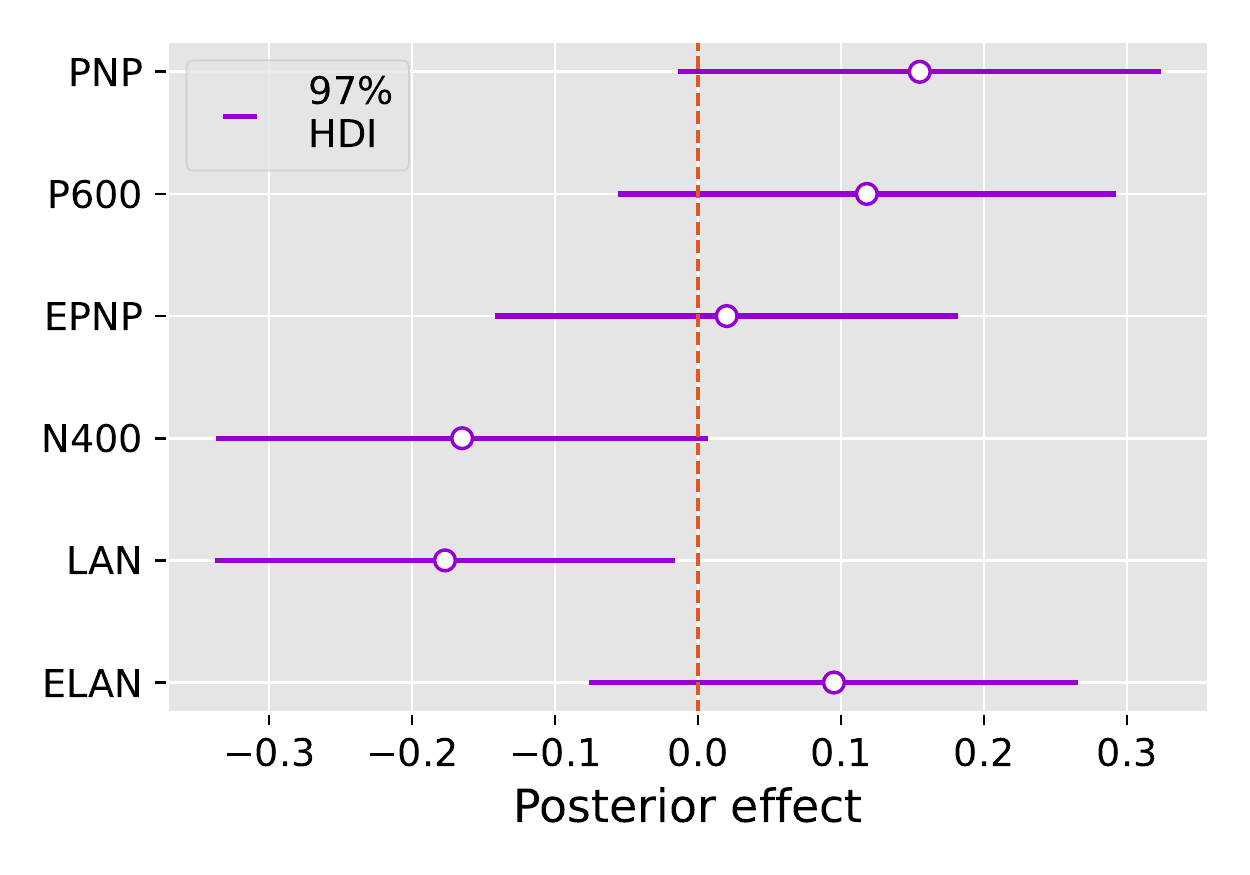}
\end{tabularx}
\caption{Posterior insights for model fitted on surprisal values from the LSTM.}
\label{fig:lstm}
\end{figure*}

\begin{figure*}[h]
\begin{tabularx}{.99\linewidth}{ 
  X X X  }
\textbf{D}&\textbf{E}&\textbf{F}\\[-3pt]
\includegraphics[width=0.32\textwidth]{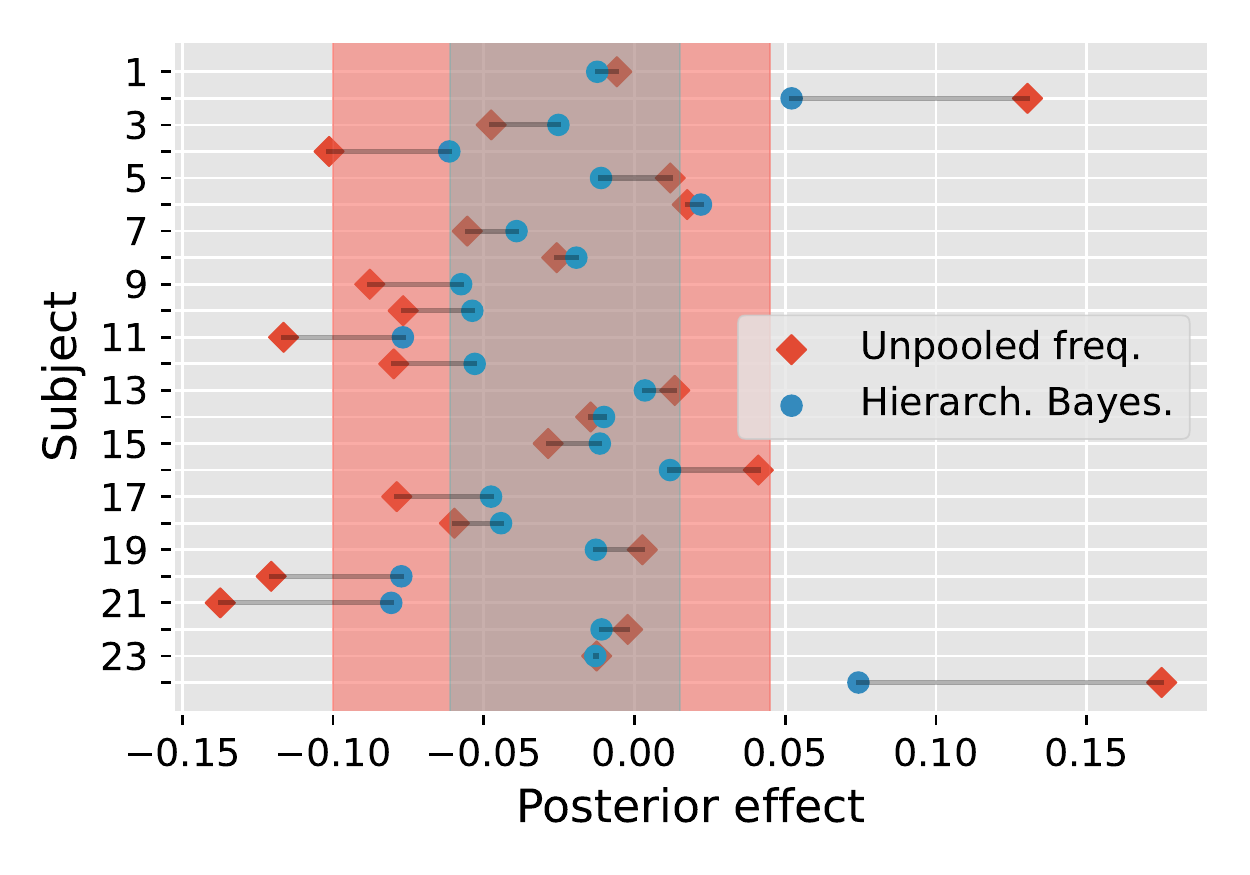}&
\includegraphics[width=0.32\textwidth]{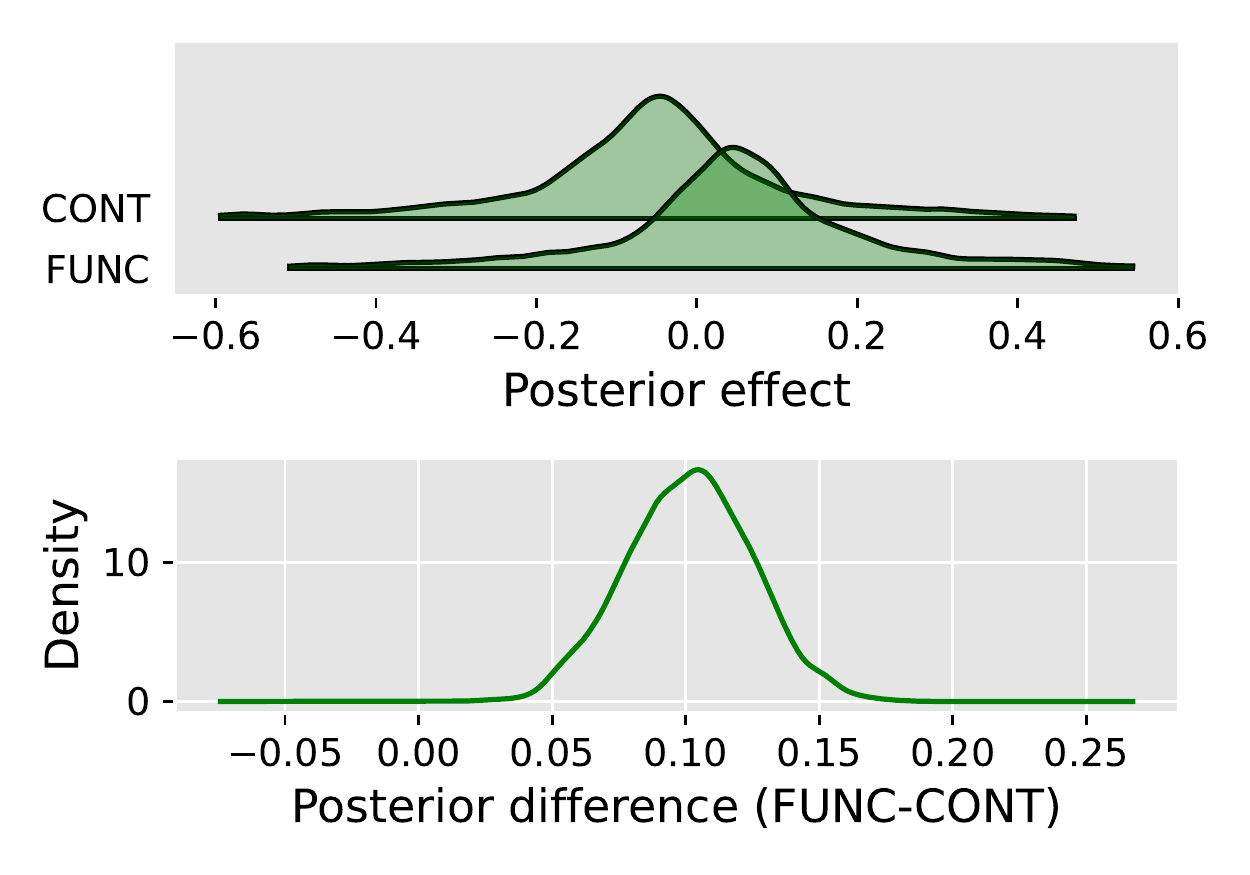}&
\includegraphics[width=0.32\textwidth]{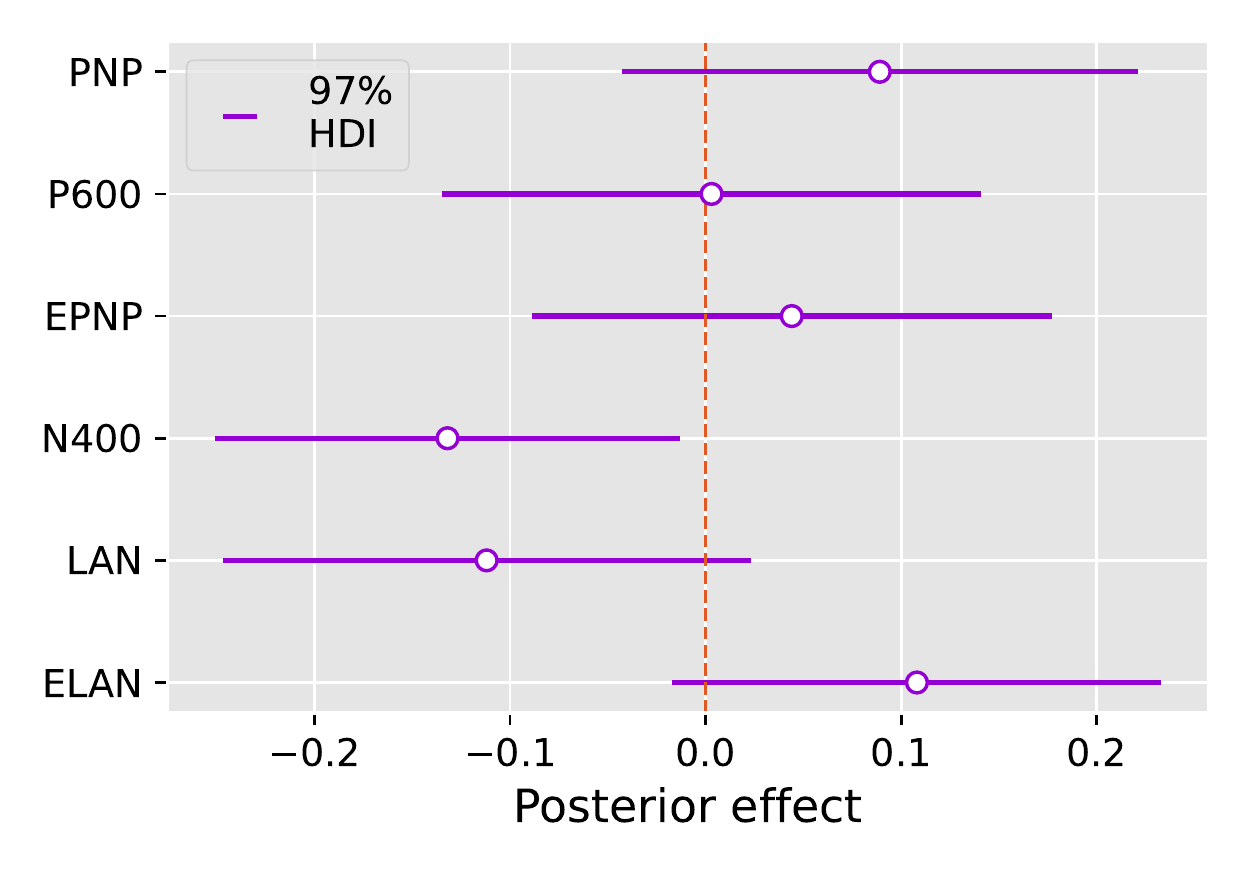}
\end{tabularx}
\caption{Posterior insights for model fitted on surprisal values from GPT-2.}
\label{fig:gpt2}
\end{figure*}

The posterior distributions for the models fitted on LSTM and GPT-2 surprisal values are illustrated in Fig. \ref{fig:lstm} and \ref{fig:gpt2} respectively. Overall, results are similar to $n$-gram effects shown in Fig. \ref{fig:results}, although smaller in magnitude for the model fitted on GPT-2 surprisal values. Moreover, by looking at the by-ERP-component posterior effect estimates for this LM (Fig.  \ref{fig:gpt2}\textbf{F}), it appears that the model may be unable to distinguish between the negativities N400 and LAN, as well as the positivities P600 and EPNP. 

\subsection{Diagnostics and Model Comparison}
\begin{figure*}[h]
    \centering
    \includegraphics[width=.42\textwidth]{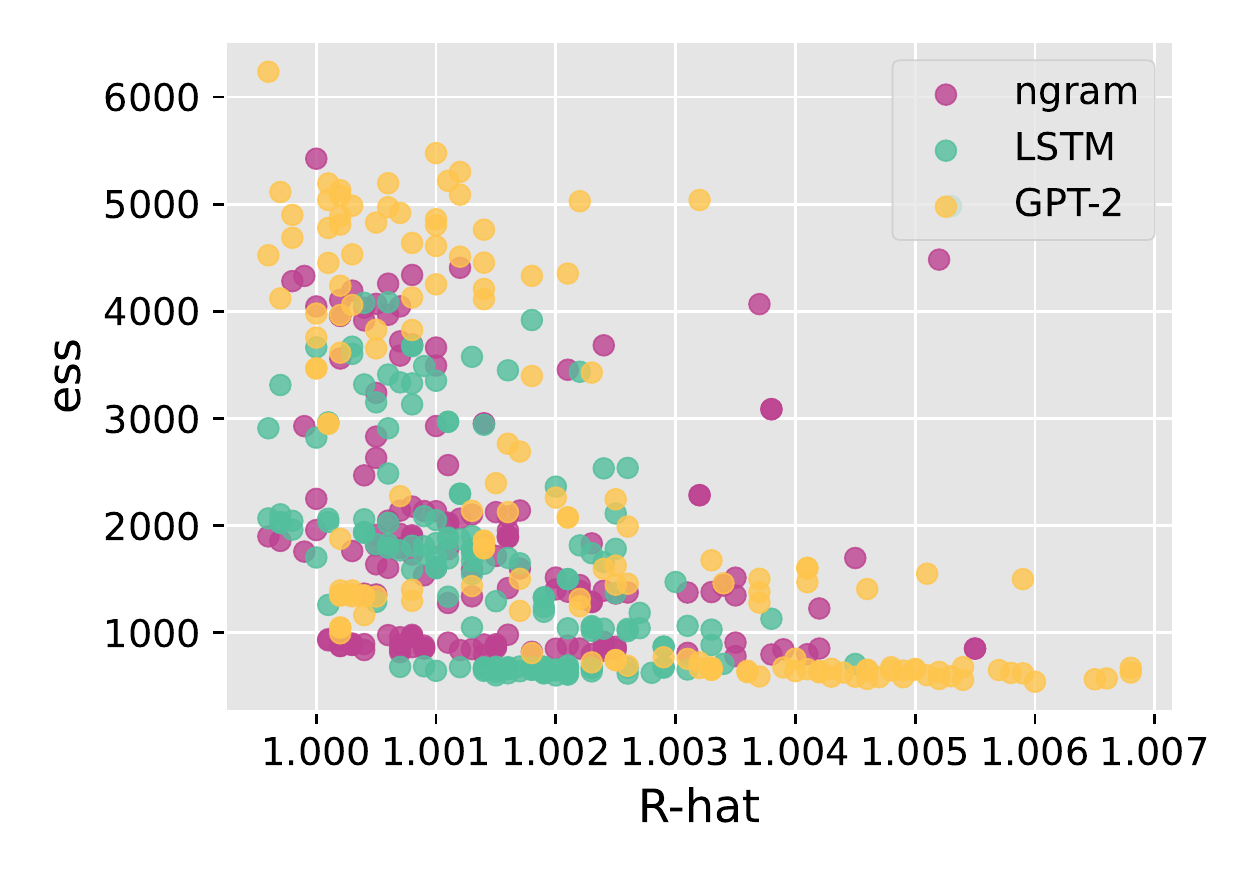}
    \caption{Diagnostic tests for the sampler; each point represents a parameter value in a chain. For all the LMs, the model reaches convergence, as shown in terms of R-hat statistics and effective sample size (ess).}
    \label{fig:diagn}
\end{figure*}

The quality of sampling is evaluated with two common diagnostic criteria: R-hat and effective sample size (ess). The former is a measure of convergence of the chains to a stationary distribution, whereas the latter corresponds to the effective number of samples in terms of autocorrelation \cite{McElreath2020, Primer}. Usually, convergence and sampling are deemed satisfactory when R-hat is below $1.05$ and ess above $20\%$ of the number of samples drawn. As shown in Fig. \ref{fig:diagn}, the model successfully converges when fitted on the surprisal values from all the three LMs investigated.

\begin{figure*}[h]
\begin{tabularx}{.99\linewidth}{ 
  X X }
\textbf{A}&\textbf{B}\\[-3pt]
\includegraphics[width=0.5\textwidth]{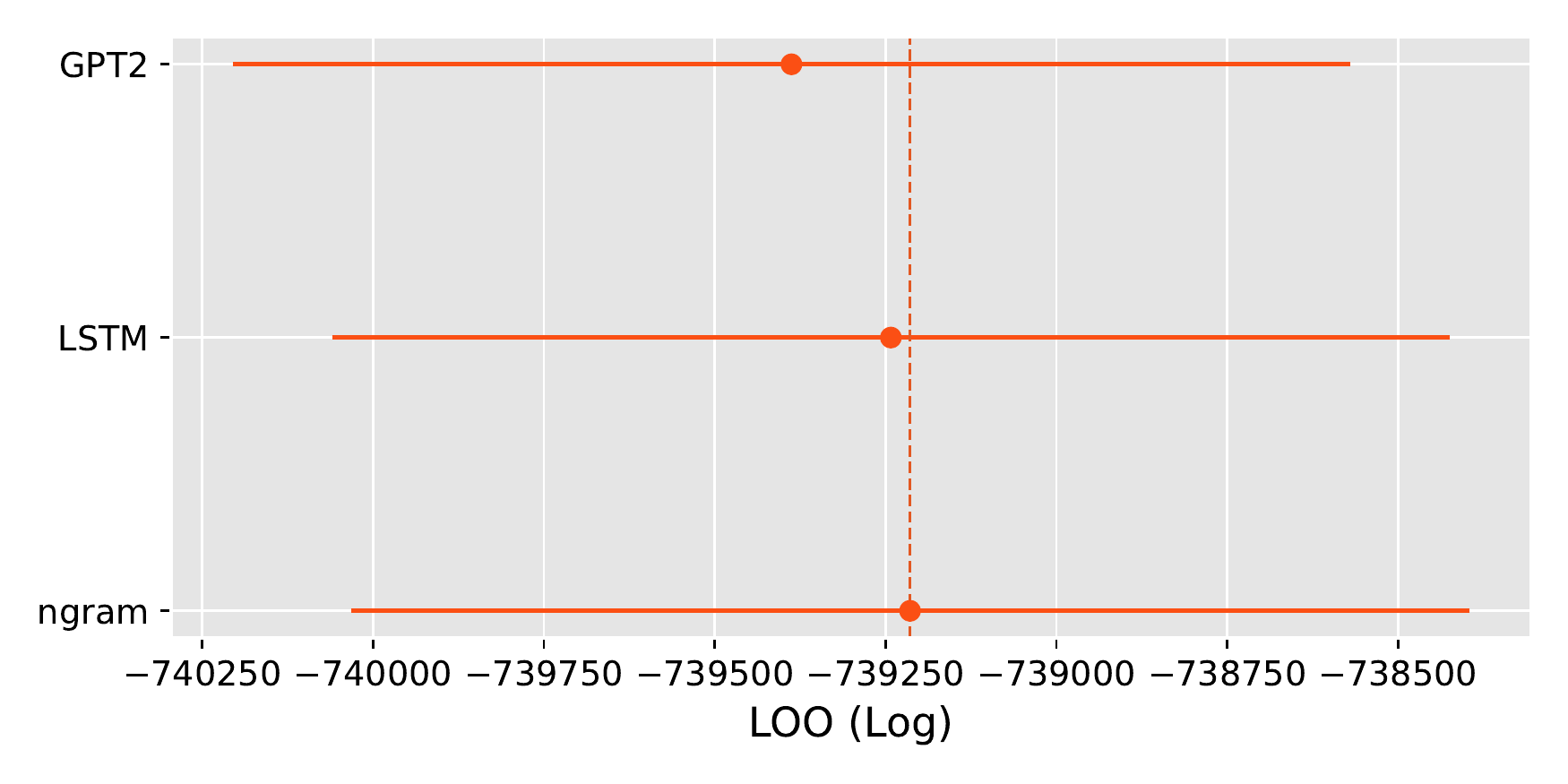}&
\includegraphics[width=0.38\textwidth]{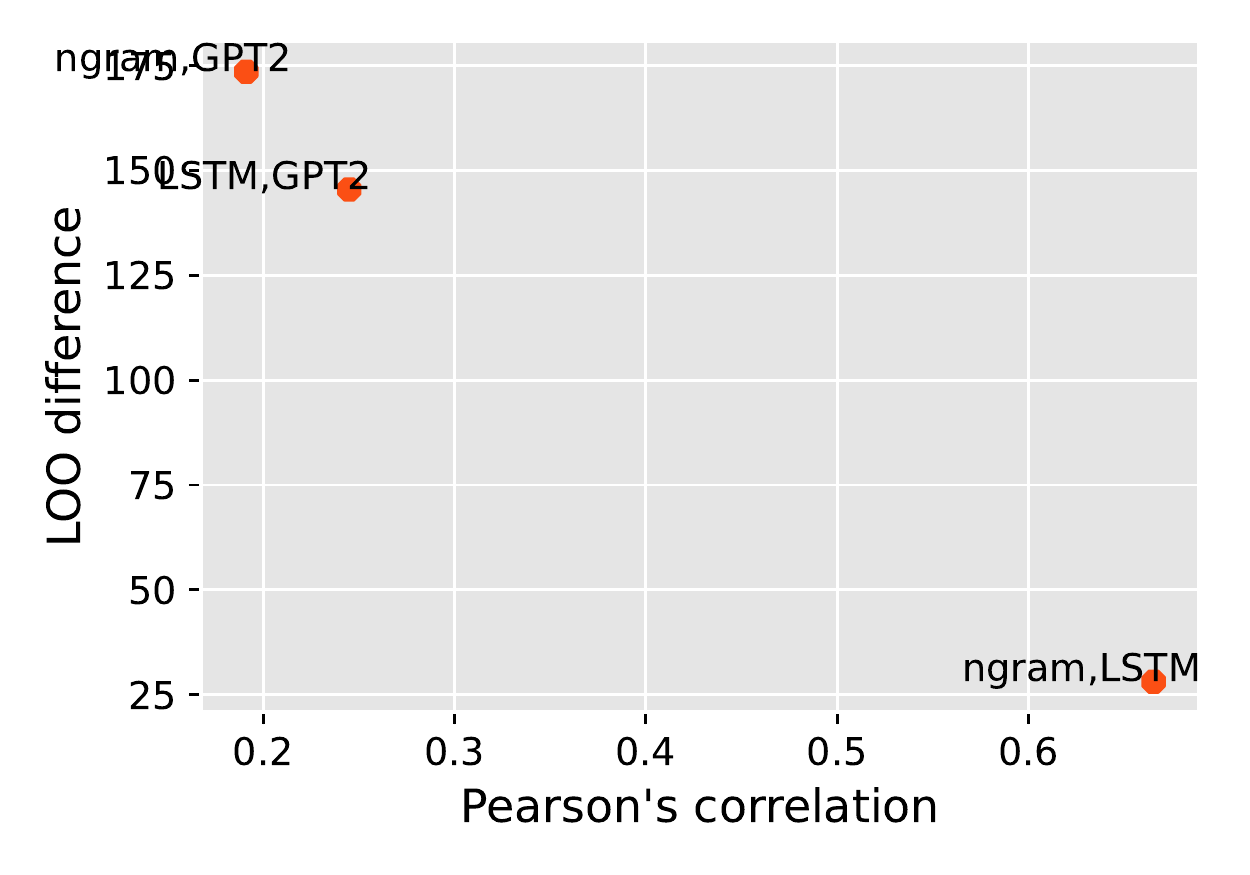}
\end{tabularx}
\caption{\textbf{A} Model comparison between models fitted on surprisal values from the three different LMs. Higher LOO indicates better out-of-sample predictive performance. It is clear that models have almost identical predictive performances. \textbf{B} LMs with more correlated surprisal estimates seem to have more similar out-of-sample predictive performances in terms of LOO.}
\label{fig:compar}
\end{figure*}

Although this work concerns statistical inference on effect magnitudes, for reference we include results of model comparison analysis based on LOO cross-validation with Pareto smoothing. Comparison results are illustrated in Fig. \ref{fig:compar}\textbf{A}: the model fitted on the three LMs show almost identical out-of-sample predictive performances. Moreover, Fig. \ref{fig:compar}\textbf{B} shows that small differences in LOO between the model fitted on different surprisal values are inversely proportional to the Pearson's correlation between the surprisal values from the three LMs, as expected.


\end{document}